\documentclass[10pt,conference]{IEEEtran}
\IEEEoverridecommandlockouts

\usepackage{cite}
\usepackage{textcomp}
\usepackage{balance}

\usepackage{longtable}
\usepackage[colorlinks=true,allcolors=blue]{hyperref}
\usepackage{amsmath,amssymb,amsfonts}
\usepackage{graphicx}
\usepackage{xcolor}
\usepackage{booktabs} 
\usepackage{multirow}
\usepackage{listings}
\usepackage{tcolorbox}

\usepackage{xcolor}
\usepackage{listings}
\usepackage{tcolorbox}
\usepackage{caption}

\definecolor{added}{RGB}{198,239,206}    
\definecolor{deleted}{RGB}{255,199,206}  
\definecolor{graycomment}{gray}{0.4}

\lstdefinestyle{diffstyle}{
    basicstyle=\ttfamily\small,
    breaklines=true,
    columns=fullflexible,
    showstringspaces=false,
    moredelim=**[is][\color{graycomment}]{@}{@}, 
    moredelim=**[is][\color{black}\bfseries\setlength{\fboxsep}{0pt}\colorbox{deleted}]{-}{-},
    moredelim=**[is][\color{black}\bfseries\setlength{\fboxsep}{0pt}\colorbox{added}]{+}{+},
}

\usepackage{relsize} 
\usepackage{cleveref}
\usepackage{soul}
\usepackage{lineno}
\usepackage{color, colortbl}
\usepackage{csquotes}
\usepackage{amssymb}
\usepackage{xspace}
\usepackage{fontawesome}
\usepackage{tikz,pifont}
\usepackage{float}

\colorlet{lightgray}{gray!30}

\sethlcolor{lightgray}

\definecolor{mygreen}{rgb}{0,0.6,0}
\definecolor{mygray}{rgb}{0.95,0.95,0.95}
\definecolor{myred}{rgb}{0.5,0,0}

\lstdefinestyle{JavaStyle} {
	backgroundcolor=\color{white},   
	commentstyle=\color{mygreen}, 
	breakatwhitespace=false,
	keywordstyle=\color{violet},
	language=Java,
	stringstyle=\color{blue},
	basicstyle=\scriptsize\ttfamily,
	showstringspaces=false }

\newcommand*{\tool}{\texttt{CARL-CCI}\@\xspace}
\newcommand*{\ccisolver}{\texttt{CCI-Solver}\@\xspace}
\newcommand*{\cforllama}{\texttt{C4RLLaMA}\@\xspace}
\newcommand*{\deepjit}{\texttt{DeepJIT}\@\xspace}

\newcommand*{\unixcoder}{\texttt{UniXcoder}\@\xspace}
\newcommand*{\codellama}{\texttt{CodeLLaMA}\@\xspace}
\newcommand*{\deepseekcoder}{\texttt{DeepSeek-Coder}\@\xspace}

\newcommand*{\qwen}{\texttt{Qwen2.5-Coder}\@\xspace}

\newcommand*{\jitdata}{\texttt{JIT\textsmaller{DATA}}\@\xspace}

\newcommand*{\ccibench}{\texttt{CCI\textsmaller{BENCH}}\@\xspace}

\newcommand*{\posthoc}{\textit{post-hoc}\@\xspace}

\newcommand{\rqone}{\textbf{RQ$_1$}: \emph{How effective is the proposed model compared to
existing baselines?}}

\newcommand{\rqtwo}{\textbf{RQ$_2$}: \emph{How does each component of \tool contribute to the final performance?}}

\usepackage{amssymb}

\newtcolorbox{shadedbox}{
	drop shadow southeast,
	breakable,
	enhanced jigsaw,
	colback=white,
	boxrule=0.80pt,
	left=0.3em,
	right=0.3em,
	top=0.1em,
	bottom=0.05em
}

\def\BibTeX{{\rm B\kern-.05em{\sc i\kern-.025em b}\kern-.08em
    T\kern-.1667em\lower.7ex\hbox{E}\kern-.125emX}}

\begin{document}

\title{Larger Is Not Always Better: Leveraging Code Evolution for Comment Inconsistency Detection}


\title{Larger Is Not Always Better: Leveraging Structured Code Diffs for Comment Inconsistency Detection}

\author{\IEEEauthorblockN{Phong Nguyen}
\IEEEauthorblockA{
\textit{Hanoi University of Science and Technology}\\
Hanoi, Vietnam\\
phong.nhv210669@sis.hust.edu.vn
}
\and
\IEEEauthorblockN{Anh M. T. Bui$^{*}$\thanks{*Corresponding author}}
\IEEEauthorblockA{
\textit{Hanoi University of Science and Technology}\\
Hanoi, Vietnam \\
anhbtm@soict.hust.edu.vn}
\and
\IEEEauthorblockN{Phuong T. Nguyen}
\IEEEauthorblockA{
\textit{University of L'Aquila}\\
L'Aquila, Italy \\
phuong.nguyen@univaq.it}
}




\maketitle

\begin{abstract}

Ensuring semantic consistency between source code and its accompanying comments is crucial for program comprehension, effective debugging, and long-term maintainability. Comment inconsistency arises when developers modify code but neglect to update the corresponding comments, potentially misleading future maintainers and introducing errors. Recent approaches to code–comment inconsistency (CCI) detection leverage Large Language Models (LLMs) and rely on capturing the semantic relationship between code changes and outdated comments. However, they often ignore the structural complexity of code evolution, including historical change activities, and introduce privacy and resource challenges.
In this paper, we propose a Just-In-Time CCI detection approach built upon the \texttt{CodeT5+} backbone. Our method decomposes code changes into ordered sequences of modification activities such as replacing, deleting, and adding to more effectively capture the correlation between these changes and the corresponding outdated comments.
Extensive experiments conducted on publicly available benchmark datasets--\texttt{JITDATA} and \texttt{CCIBENCH}--demonstrate that our proposed approach outperforms recent state-of-the-art models by up to 13.54\% in F1-Score and achieves an improvement ranging from 4.18\% to 10.94\% over fine-tuned LLMs including \texttt{DeepSeek-Coder}, \texttt{CodeLlama} and \texttt{Qwen2.5-Coder}.


\end{abstract}


	\section{Introduction}
	\label{sec:intro}
\begin{figure}[t]
    \centering
        \includegraphics[width=0.9\linewidth]{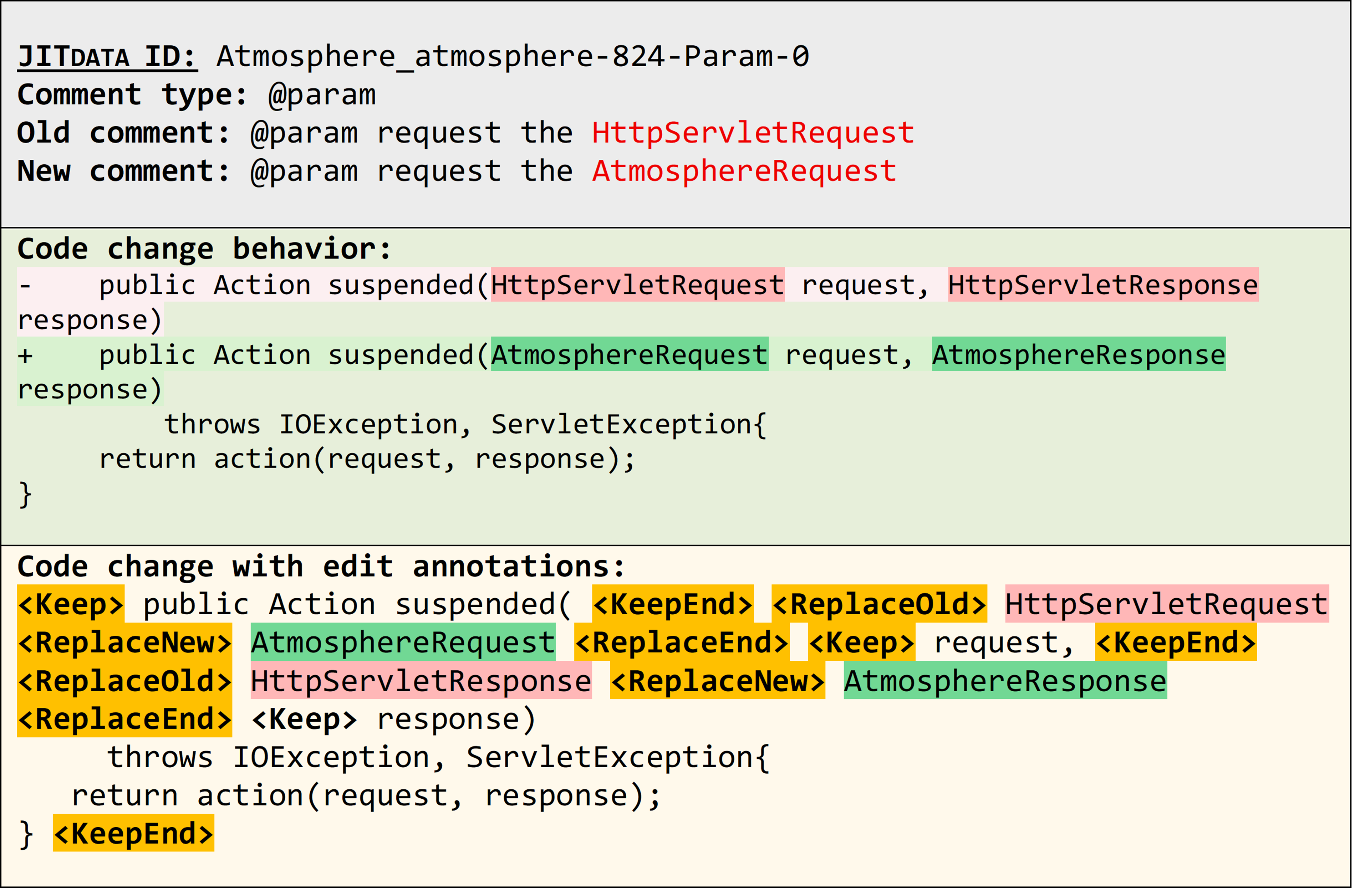}
    \caption{A CCI example in \jitdata~\cite{Panthaplackel2020DeepJI}.}
    \label{fig:sample}
\end{figure}

	Code comments are essential documentation artifacts in software development, enabling developers to communicate a program’s functionality, design rationale, and intended usage~\cite{Padioleau2009ListeningTP}. Well-maintained comments have been shown to enhance developers’ comprehension of complex systems and facilitate more efficient software maintenance.
However, as software systems evolve through continuous modifications and feature extensions, developers often fail to update corresponding comments, leading to code–comment inconsistency (CCI), where comments no longer align with the underlying implementation~\cite{Wen2019ALE, Fluri2009AnalyzingTC, Tan2007icommentBO}. Such inconsistencies can mislead future maintainers, hinder debugging, and ultimately increase the likelihood of software defects~\cite{Jiang2006ExaminingTE, Ibrahim2012OnTR}. For instance, Figure\ref{fig:sample} illustrates a
CCI scenario in \jitdata where a method was
modified to handle AtmosphereRequest objects, yet the old comment incorrectly claims it works with HttpServletRequest.




Early studies on CCI detection focused on identifying inconsistencies between static code and comments by analyzing a fixed snapshot of the repository~\cite{Corazza2016CoherenceOC,Cimasa2019WordEF}. However, such approaches are limited in handling the inherently evolving nature of software systems, where comments may become outdated only after subsequent code modifications. To better cope with this dynamic setting, research has gradually shifted toward Just-in-time (JIT) detection, where models operate over sequences of commits and leverage code changes to identify potential mismatches ~\cite{Rong2025CodeCI, Zhong2025CCISolverED,Panthaplackel2020DeepJI}. A wide range of techniques have been explored for this task, ranging from heuristic-based approaches~\cite{Ratol2017DetectingFC} to deep learning models~\cite{Liu2018AutomaticDO,Allamanis2017ASO, Panthaplackel2020DeepJI,Liu2023JustInTimeOC, Xu2023DataQM}, and more recently LLM-based methods~\cite{zhang2024detecting,Rong2025CodeCI}. Despite these advances, most existing approaches still treat code diffs as flattened token sequences, which obscures the distinction between added and deleted regions. As a result, current models struggle to capture the fine-grained correlation links between specific edits and the comment, limiting their ability to reason about semantic drift introduced by code changes.

In this paper, we propose \tool, a just-in-time approach to CCI detection, drawing on different code-change activities to infer the correlation between code modifications and old comments so as to identify inconsistencies.
We analyze code diffs and annotate each edit with its corresponding activity including \texttt{<replace>}, \texttt{<add>}, or \texttt{<delete>}. Rather than treating changes as a monolithic token sequence, we tag each token span with its activity label and feed the resulting sequence, together with the comment, into a cross-encoder. This design highlights correlation connections between specific edits and the comment via the explicit change-activity signal. 
The upper section of Figure~\ref{fig:sample} depicts a CCI example; diffs are highlighted with red for deletions and green for additions. 
In the lower part, activity labels are applied to token spans and highlighted in yellow.
Using the \texttt{CodeT5+} model as backbone, we propose a label-aware contrastive learning framework that more effectively captures the semantics of code diffs and their comments. In contrast to self-supervised, monolithic contrastive objectives, our approach uses ground-truth labels to attract same-label samples and repel different-label ones, producing a more organized and discriminative representation space for distinguishing inconsistent from consistent comments.

The main contributions of our work are as follows.

\begin{itemize}
	\item We redesign code-diff representations to embed an activity label within each token span, allowing us to probe the correlation relationship between edits and comments.
    \item We introduced \tool, a {\it just-in-time} approach to code comment inconsistency detection that builds on top of the \texttt{CodeT5+} model and leverages label-aware contrastive learning to enhance representations of code diffs and comments.
	\item By conducting extensive experiments on two popular benchmarks, we observed consistent improvements over previous approaches; moreover, relative to fine-tuned LLMs, our method delivers stronger performance for detecting inconsistent comments.
	\item We published the replication package, including code and data to facilitate future research~\cite{CARLCCI26}.
\end{itemize}


\noindent
\textbf{Structure.}
In Section~\ref{sec:rw}, we review
 the related work. Section~\ref{sec:ProposedApproach} explains in detail the proposed
 approach. The empirical evaluation to study the performance of \tool is presented in Section~\ref{sec:settings}. Afterward, in Section~\ref{sec:result}, we report and analyze the experimental results. Finally,
 Section~\ref{sec:conclusion} sketches future work, and concludes the paper.

	\section{Related Work}
	\label{sec:rw}

Research on code--comment inconsistency (CCI) detection spans two primary settings: \emph{post-hoc} and \emph{Just-in-time} (JIT).

Early \posthoc studies focused on identifying inconsistencies in existing repositories by analyzing static code–comment pairs.
These efforts mainly relied on rule-based heuristics (e.g., lexical overlap~\cite{Ratol2017DetectingFC}, identifier matching~\cite{Zhou2017AnalyzingAD}) or shallow learning techniques such as similarity-based models using \texttt{TF}–\texttt{IDF} or word embeddings~\cite{Corazza2016CoherenceOC, Cimasa2019WordEF}.
However, these \emph{post-hoc} approaches are limited by their inability to account for the evolving nature of code.

Motivated by practical needs, research has shifted toward JIT detection, which leverages commit-level code changes to identify inconsistencies at the moment they are introduced. Panthaplackel et al.\ introduced the \jitdata dataset and the \deepjit framework~\cite{Panthaplackel2020DeepJI}, establishing the first large-scale benchmark for commit-level CCI detection. Subsequent work explored stronger architectures, including transformer-based models~\cite{Steiner2022CodeCI, Xu2024CodeCI} and LLM-based systems such as \cforllama~\cite{Rong2025CodeCI}, which support both detection and comment rectification.
However, \jitdata was constructed using heuristic rules, resulting in substantial label noise that undermines the reliability of model evaluation and limits the consistency of learned representations.


To address this issue, Zhong et al.\ constructed \ccibench \cite{Zhong2025CCISolverED}, a higher-quality benchmark created via syntactic cleaning and ensemble validation using state-of-the-art LLMs including \texttt{GPT}-\texttt{4o} \cite{OpenAI2024GPT4o}, \texttt{Claude}\texttt{3.5}-\texttt{Sonnet} \cite{Anthropic2024Claude35}, and \texttt{LLaMA~3.1}-\texttt{405B} \cite{Meta2024Llama3}. CCIBench mitigates noisy labels and offers a more dependable basis for both detection and repair tasks.

Despite these developments, most existing models still process code diffs as single flattened sequences, without explicitly distinguishing added from deleted code segments\cite{Panthaplackel2020DeepJI, Dau2023DocCheckerBC, Rong2025CodeCI}.
This limitation prevents them from capturing the correlation relationships between specific edits and their corresponding comments, thereby hindering fine-grained inconsistency detection.

	\section{Proposed Approach}
	\label{sec:ProposedApproach}

\begin{figure*}[t]
    \centering
    \includegraphics[width=\textwidth]{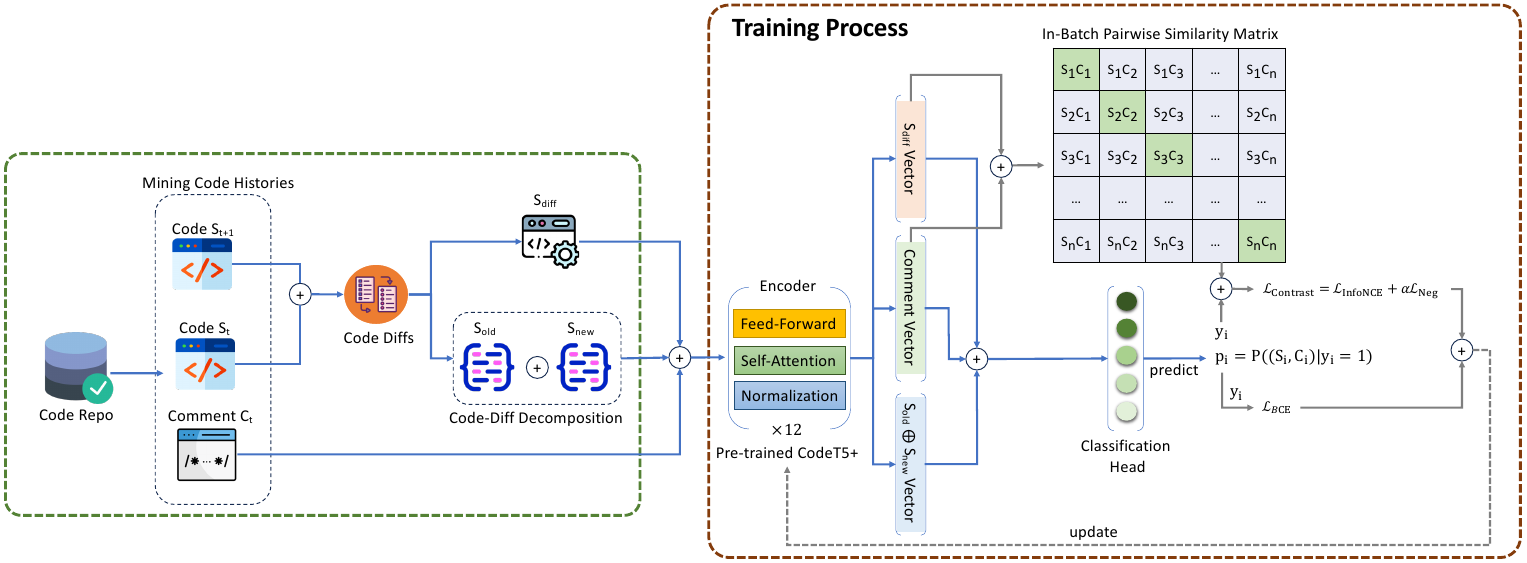}
    \caption{The overall architecture of our proposed approach.}
    \label{fig:strategy2}
\end{figure*}

In this section, we introduce \tool--Contrastive Activity-aware Representation Learning for CCI detection. Figure~\ref{fig:strategy2} outlines two components: (i) a pre-processing module that extracts code diffs and re-formats them with edit labels, and (ii) a cross-encoder that jointly encodes the concatenated code-diff/comment pair to capture fine-grained correlations between specific edits and comments, going beyond global similarity of prior works~\cite{Panthaplackel2020DeepJI,Rong2025CodeCI}. To make these interactions discriminative, we employ label-aware contrastive learning, which pulls together consistent pairs and pushes apart inconsistent ones, shaping a structured embedding space for consistency classification. 


\subsection{Activity-Labeled Code-Diff Representation} 
Given a code snippet $\mathcal{S}$ and its comment $\mathcal{C}$, we define a function $\mathcal{I}(\cdot,\cdot)$ as follows.
\[
\mathcal{I}(\mathcal{S},\mathcal{C}) = 
\begin{cases}
0 & \text{if } \mathcal{C} \text{ accurately captures the semantics of } \mathcal{S},\\
1 & \text{otherwise.}
\end{cases}
\]




In line with prior work~\cite{Liu2018AutomaticDO,Panthaplackel2020DeepJI,Liu2023JustInTimeOC,Rong2025CodeCI}, we assume that comment updates indicate a preceding code change would have caused inconsistency. 
Accordingly, the pre-change pair is taken as consistent, i.e., $\mathcal{I}(\mathcal{S}_t,\mathcal{C}_t)=0$.
After a code update, $\mathcal{S}_t \rightarrow \mathcal{S}_{t+1}$, the task is to judge whether the unchanged comment $\mathcal{C}_t$ has become inconsistent. 
Given $\mathcal{S}_{\text{diff}}$ denotes the code diffs between $\mathcal{S}_t$ and $\mathcal{S}_{t+1}$, the CCI detection is then formulated as follows.
\[
\mathcal{I}(\mathcal{S}_{\text{diff}}, \mathcal{C}_t) =
\begin{cases}
1 & \text{if } \mathcal{C}_t \text{ does not match the semantics of } \mathcal{S}_{t+1},\\
0 & \text{otherwise.}
\end{cases}
\]


We reformulate code-diff sequences using four edit actions—\textsc{Add}, \textsc{Del}, \textsc{Keep}, and \textsc{Replace}—and insert tags to mark each edit span.
An example of reformulated code diffs is depicted in Figure~\ref{fig:sample}.
A code-diff sequence $\mathcal{S}_{\text{diff}}$ is now represented as a mixture of four kinds of edit span as follows.
\begin{itemize}
    \item \texttt{<Add>} tokens \texttt{<EndAdd>}
    \item \texttt{<Del>} tokens \texttt{<EndDel>}
    \item \texttt{<Keep>} tokens \texttt{<EndKeep>}
    \item \texttt{<ReplaceOld>} tokens \texttt{<ReplaceNew>} tokens \texttt{<EndReplace>}
\end{itemize}

In order to capture the correlation relationship between different kinds of code diffs and comments, we decompose $\mathcal{S}_{\text{diff}}$ into two parts: $\mathcal{S}_{\text{diff}} = \mathcal{S}_{\text{old}} \bigcup \mathcal{S}_{\text{new}} \bigcup \mathcal{S}_{\text{unchanged}}$ where:
\begin{itemize}
    \item $\mathcal{S}_{\text{old}} = \{\text{Del-tokens} \cup \text{ReplaceOld-tokens}\}$
    \item $\mathcal{S}_{\text{new}} = \{\text{Add-tokens} \cup \text{ReplaceNew-tokens}\}$
    \item $\mathcal{S}_{\text{unchanged}} = \{\text{Keep-tokens}\}$
\end{itemize}






Under the assumption that $\mathcal{I}(\mathcal{S}_t, \mathcal{C}_t)=0$, we reformulate the detection objective as:
\[
\mathcal{I}(\mathcal{S}_{\text{diff}}, C_t) =
\begin{cases}
0 & \text{if } f(\mathcal{S}_{\text{old}}, C_t) = 0 \ \wedge\ f(\mathcal{S}_{\text{new}}, C_t) = 0,\\
1 & \text{if } f(\mathcal{S}_{\text{old}}, C_t) = 1 \ \wedge\ f(\mathcal{S}_{\text{new}}, C_t) = 0.
\end{cases}
\]
where $f(\cdot,\cdot)$ denotes a learned relevance function measuring semantic alignment between code edits and the comment. 
Decomposing the code-diff sequence guides the model toward local, fine-grained edit–comment relations, yielding more interpretable and resilient inconsistency detection than holistic formulations.

\subsection{Contrastive and Activity-Aware Representation Learning for CCI}


\subsubsection{Cross-Encoding Code Diffs and Comments}
We use \texttt{CodeT5+} as the backbone, employing its encoder to jointly encode code-diff/comment pairs. For a given code diff and its associated comment $(\mathcal{S}_{\text{diff}},\mathcal{C})$, the encoder input is constructed as follows.
\[
x = \mathcal{S}_{\text{old}} \oplus \mathcal{S}_{\text{new}} \oplus \mathcal{C} \oplus \mathcal{S}_{\text{diff}}
\]
Action-tagged edit spans allow the cross-encoder to capture edit–comment correlations, letting it focus attention on the edits most pertinent to the comment.

By concatenating the vector representations of the code diff and the comment, we obtain a joint input to the classifier that predicts inconsistency.
We fine-tune the encoder and classification head end-to-end using the binary cross-entropy $\mathcal{L}_{BCE}$ (as in Equation~\ref{eq:ce}).
\begin{equation}
    \mathcal{L}_{BCE}(\mathcal{B}) = -\frac{1}{|\mathcal{B}|}\sum\limits_{i \in \mathcal{B}}(y_ilog(p_i)+(1-y_i)log(1-p_i))
    \label{eq:ce}
\end{equation}
where $\mathcal{B}$ is the training batch, $|\mathcal{B}|$ denotes the number of training samples in batch, $y_i$ represents the ground-truth label of sample $i$ ($y_i \in \{0,1\}$) and $p_i$ is the model’s predicted inconsistency probability for that sample.

\subsubsection{Label-Aware Contrastive Learning}
To better encode the semantics of consistent and inconsistent pairs, we adopt supervised (label-aware) contrastive learning: same-label pairs are aligned, different-label pairs are repelled, producing a more separable representation. 
Concretely, we extend \texttt{InfoNCE}~\cite{Radford2021LearningTV}: (i) for $\text{label}=0$ (consistent pairs), we use \texttt{InfoNCE} to bring the matched diff–comment pair closer; (ii) for $\text{label} = 1$ (inconsistent pairs), we incorporate a negative-push term~\cite{Chen2022DualCL} to drive mismatched pairs apart.

Given a training batch $\mathcal{B}$, we compute a pairwise similarity between every comment $c_i$ and code diff $s_j$ as in Equation~\ref{eq:pw}.

\begin{equation}
sim_{ij} = \frac{\mathbf{c}_i \cdot \mathbf{s}_j}{\tau},
\qquad
p_{ij} = \frac{\exp(sim_{ij})}{\sum\limits_{k \in \mathcal{B}} \exp(sim_{ik})}.
\label{eq:pw}
\end{equation}
where $sim_{ij}$ is the similarity between the $i$-th comment and the $j$-th code diff, $p_{ij}$ is the row-wise softmax probability and $\tau$ is the temperature in contrastive learning.

The training batch $\mathcal{B}$ is divided into two subsets, let $\mathcal{B}_C = \{\text{pair}_i | \text{label}_i = 0\}$ and $\mathcal{B}_{I} = \{\text{pair}_j|\text{label}_j = 1\}$. Two supervised contrastive loss components are formulated as in Equations~\ref{eq:lossNCE} and~\ref{eq:lossNeg}.

\begin{equation}
L_{\text{InfoNCE}}
= -\frac{1}{|\mathcal{B}_C|} \sum_{i\in \mathcal{B}_C} \log p_{ii}
\label{eq:lossNCE}
\end{equation}

\begin{equation}
L_{\text{neg}}
= -\frac{1}{|\mathcal{B}_I|} \sum_{j\in \mathcal{B}_I} \log(1 - p_{jj})
\label{eq:lossNeg}
\end{equation}

The contrastive loss, composed of these two components, is defined as follows.
\begin{equation}
L_{\text{contrast}}
= L_{\text{InfoNCE}} + \alpha L_{\text{neg}},
\qquad
\alpha \ge 0,
\label{eq:contrastive}
\end{equation}
where $\alpha$ is a weighting coefficient that controls the contribution of the inconsistent-pair term. The overall training objective is then given in Equation~\ref{eq:total-loss}.
\begin{equation}
L_{\text{total}}
= L_{\text{CE}} + \lambda L_{\text{contrast}},
\qquad
\lambda \ge 0.
\label{eq:total-loss}
\end{equation}
where $\lambda$ sets the proportion of the contrastive component in the total loss.






	\section{Evaluation} 
	\label{sec:settings}

\subsection{Research Question}
Our study explores the following research questions.

\smallskip
\noindent
$\triangleright$ \rqone~We evaluate the effectiveness of \tool in comparison with state-of-the-art approaches. Specifically, we select three recent baseline models, including:
\begin{itemize}
    \item \deepjit~\cite{Panthaplackel2020DeepJI} employs deep learning architectures to capture information from code diffs and comments.

    \item \cforllama~\cite{Rong2025CodeCI} fine-tunes \codellama and leverages chain-of-thought reasoning
    to improve both CCI detection and comment rectification.

    \item \ccisolver~\cite{Zhong2025CCISolverED} uses the pre-trained \unixcoder to detect inconsistencies.
\end{itemize}

Additionally, we fine-tune several code-oriented models, including \deepseekcoder~\cite{DeepSeekAI2024DeepSeekCoderV2BT}, \qwen~\cite{Hui2024Qwen25CoderTR}, and \codellama~\cite{Rozire2023CodeLO}.  
Specifically, we use the \texttt{deepseek-coder-6.7b-instruct}, \texttt{Qwen2.5-Coder-7B-Instruct}, and \texttt{CodeLlama-7b}-\texttt{Instruct-hf} variants. 

\smallskip
\noindent
$\triangleright$ \rqtwo~
We evaluate the contribution of each component by ablating key modules--such as the activity-labeled code diff representation and the label-aware contrastive loss--and measuring the resulting performance changes.

\subsection{Benchmark Datasets}
\label{sec:benhmark-datasets}
We conduct experiments on two benchmark datasets widely used in CCI research: \jitdata~\cite{Panthaplackel2020DeepJI} and \ccibench~\cite{Zhong2025CCISolverED}.  
\jitdata contains 40,688 code-diff/comment pairs collected from 1,518 open-source Java projects, while \ccibench is a cleaned and higher-quality derivative of this dataset.  
Following prior work, all experiments will be performed on the full test set and on the validated test set which is constructed by manually selecting 300 verified instances from the test set to support more reliable evaluation. Table~\ref{tab:dataStat} summarizes the dataset statistics of two benchmarks.


\begin{table}[t]
\centering
\caption{Statistics of the \jitdata and \ccibench datasets.}
\label{tab:dataStat}
\begin{tabular}{llcccc}
\toprule
\textbf{Dataset} & \textbf{Comment Type} & \textbf{Train} & \textbf{Validation} & \textbf{Test} & \textbf{Total} \\
\midrule
\multirow{4}{*}{\textit{JIT\textsmaller{DATA}}}
  & @return  & 15{,}950 & 1{,}790 & 1{,}840 & 19{,}580 \\
  & @param   & 8{,}640  & 932     & 1{,}038 & 10{,}610 \\
  & Summary  & 8{,}398  & 1{,}034 & 1{,}066 & 10{,}498 \\
  & \textbf{All} & \textbf{32{,}988} & \textbf{3{,}756} & \textbf{3{,}944} & \textbf{40{,}688} \\
\midrule
\multirow{4}{*}{\textit{CCI\textsmaller{BENCH}}}
  & @return  & 9{,}188  & 1{,}092 & 1{,}088 & 11{,}368 \\
  & @param   & 6{,}004  & 620     & 646     & 7{,}270  \\
  & Summary  & 2{,}970  & 356     & 396     & 3{,}722  \\
  & \textbf{All} & \textbf{18{,}162} & \textbf{2{,}068} & \textbf{2{,}130} & \textbf{22{,}360} \\
\bottomrule
\end{tabular}
\end{table}

\subsection{Evaluation Metrics}
\label{sec:metrics}
Following prior work, we evaluate performance using \texttt{Accuracy}, \texttt{Precision}, \texttt{Recall}, and \texttt{F1-Score} as in prior work~\cite{Panthaplackel2020DeepJI,Rong2025CodeCI, Zhong2025CCISolverED}. \texttt{Accuracy} reflects overall correctness, \texttt{Precision} and \texttt{Recall} measure correctness and coverage of positive predictions, and \texttt{F1} is their harmonic mean.

\subsection{Implementation Details}
\label{sec:setup}

\textsc{\tool} relies on the encoder of pre-trained \texttt{CodeT5p-220M} model. 
Our model is trained with a batch size of 32 and a learning rate of \(1\times10^{-5}\), using a contrastive temperature of \(\tau = 0.08\). Training continues for up to 20 epochs with early stopping based on the \texttt{F1}. 

Fine-tuning the selected LLMs is performed using \texttt{QLoRA} with 4-bit NF4 quantization, where only low-rank adapter weights are updated while the base model remains frozen.
We apply a \texttt{LoRA} configuration with rank \texttt{r = 32}, scaling factor \texttt{lora\_alpha = 16}, dropout \texttt{lora\_dropout = 0.1}, and adapters inserted into \texttt{q\_proj}, \texttt{k\_proj}, \texttt{v\_proj}, \texttt{o\_proj}, \texttt{gate\_proj}, \texttt{up\_proj}, and \texttt{down\_proj}. 
All the selected LLMs are fine-tuned and evaluated on an NVIDIA H100 GPU, whereas our proposed model is trained and tested on a more modest NVIDIA T4 GPU.





	%
	\section{Results and Discussion}
	\label{sec:result}
\begin{table*}[t]
\centering
\caption{Performance of \tool comparing to baseline models on \jitdata and \ccibench.}
\small
\begin{tabular}{l>{\raggedright\arraybackslash}p{2.8cm}ccccccccc}

\toprule
\textbf{Dataset} & \textbf{Approach} &
\multicolumn{4}{c}{\textbf{Full Test Set}} &
\multicolumn{4}{c}{\textbf{Validated Test Set}} \\
\cmidrule(lr){3-6} \cmidrule(lr){7-10}
 & & \textbf{Accuracy} & \textbf{Precision} & \textbf{Recall} & \textbf{F1} 
 & \textbf{Accuracy} & \textbf{Precision} & \textbf{Recall} & \textbf{F1} \\
\midrule
\multirow{7}{*}{\jitdata} 
 & \texttt{DeepJIT}~\cite{Panthaplackel2020DeepJI}   & 82.15 & 82.85 & 81.08 & 81.96 & 89.00 & 89.26 & 88.67 & 88.96 \\
 & \texttt{C4RLLama}~\cite{Rong2025CodeCI}          & 86.24 & 86.20 & 84.30 & 85.24 & 87.82 & 89.67 & 85.33 & 87.45 \\
 & \texttt{CCISolver}~\cite{Zhong2025CCISolverED}         & 86.48 & 87.24 & 85.17 & 86.19 & 88.74 & 87.67 & 88.67 & 88.17 \\
 \cmidrule(lr){2-2} \cmidrule(lr){3-6} \cmidrule(lr){7-10}
 & \texttt{DeepSeek-Coder}~\cite{DeepSeekAI2024DeepSeekCoderV2BT} & 81.33 & 78.34 & 86.61 & 82.27 & 82.00 & 79.63 & 86.00 & 82.69 \\
 & \texttt{Qwen2.5-Coder}~\cite{Hui2024Qwen25CoderTR}      & 86.10 & 88.91 & 82.50 & 85.59 & 87.67 & 91.85 & 82.67 & 87.02 \\
 & \texttt{CodeLlama}~\cite{Rozire2023CodeLO}      & 79.79 & 74.42 & {\bf90.77} & 81.79 & 80.00 & 74.45 & 91.33 & 82.04 \\
 \cmidrule(lr){2-2} \cmidrule(lr){3-6} \cmidrule(lr){7-10}
 & \tool     & \textbf{91.18} & \textbf{94.31} & 89.67 & \textbf{90.89} & \textbf{93.67} & \textbf{95.17} & \textbf{92.00} & \textbf{93.56} \\
\midrule
\multirow{7}{*}{\ccibench} 
 & \texttt{DeepJIT}~\cite{Panthaplackel2020DeepJI}   & 87.32 & 88.41 & 85.92 & 87.15 & 89.67 & 89.40 & 90.00 & 89.70 \\
 & \texttt{C4RLLama}~\cite{Rong2025CodeCI}          & 89.16 & 88.24 & 89.10 & 88.67 & 87.82 & 92.67 & 88.67 & 90.63 \\
 & \texttt{CCISolver}~\cite{Zhong2025CCISolverED}         & 89.92 & 89.84 & 89.25 & 89.54 & 90.95 & 92.67 & 91.74 & 92.20 \\
  \cmidrule(lr){2-2} \cmidrule(lr){3-6} \cmidrule(lr){7-10}
 & \texttt{DeepSeek-Coder}~\cite{DeepSeekAI2024DeepSeekCoderV2BT}  & 83.99 & 79.48 & 91.64 & 85.12 & 84.67 & 80.23 & 92.00 & 85.71 \\
 & \texttt{Qwen2.5-Coder}~\cite{Hui2024Qwen25CoderTR}       & 88.87 & 91.48 & 85.72 & 88.51 & 90.33 & 92.91 & 87.33 & 90.03 \\
 & \texttt{CodeLlama}~\cite{Rozire2023CodeLO}       & 83.01 & 77.74 & 92.48 & 84.48 & 83.67 & 77.90 & 94.00 & 85.20 \\
  \cmidrule(lr){2-2} \cmidrule(lr){3-6} \cmidrule(lr){7-10}
 & \texttt{CARL-CCI} & \textbf{93.43} & \textbf{94.01} & \textbf{92.77} & \textbf{93.38} & \textbf{95.67} & \textbf{96.60} & \textbf{94.77} & \textbf{95.62} \\
\bottomrule
\end{tabular}
\label{tab:jit_ccibench_full}
\end{table*}

\begin{table*}[t]
\centering
\caption{Ablation Study on \jitdata and \ccibench.}
\small
\begin{tabular}{l l c cccc cccc}
\toprule
\textbf{Dataset} & \textbf{Model / Setting} &
\multicolumn{4}{c}{\textbf{Full Test Set}} &
\multicolumn{4}{c}{\textbf{Validated Test Set}} \\
\cmidrule(lr){3-6} \cmidrule(lr){7-10}
& &
\textbf{Accuracy} & \textbf{Precision} & \textbf{Recall} & \textbf{F1} &
\textbf{Accuracy} & \textbf{Precision} & \textbf{Recall} & \textbf{F1} \\
\midrule
\multirow{4}{*}{\jitdata}
 & \tool$_{\text{w/o AL and CL}}$
 & 87.55 & 88.11 & 86.82 & 87.46
 & 89.67 & 90.48 & 88.67 & 89.56 \\
 & \tool$_{\text{w/o AL}}$
 & 88.06 & 89.81 & 85.85 & 87.79
 & 88.67 & 89.73 & 87.33 & 88.51 \\
 & \tool$_{\text{w/o CL}}$
 & 91.08 & 93.94 & 88.03 & 90.74
 & 93.33  & 95.14  & 91.33  & 93.20  \\
 & \tool
 & \textbf{91.18} & \textbf{94.31} & \textbf{89.67} & \textbf{90.89}
 & \textbf{93.67}  & \textbf{95.17}  & \textbf{92.00}  & \textbf{93.56}  \\
\midrule
 \multirow{4}{*}{\ccibench}
 & \tool$_{\text{w/o AL and CL}}$
 & 89.34 & 88.44 & 90.52 & 89.47
 & 89.33 & 88.31 & 90.67 & 89.47 \\
 & \tool$_{\text{w/o AL}}$
 & 90.19 & 89.85 & 90.61 & 90.23
 & 91.00 & 89.68 & 92.67 & 91.15 \\
 & \tool$_{\text{w/o CL}}$
 & 93.10 & 93.55 & 92.58 & 93.06
 & 94.00 & 93.42 & 94.67 & 94.04 \\
 & \tool
 & \textbf{93.43} & \textbf{94.01} & \textbf{92.77} & \textbf{93.38}
 & \textbf{95.67} & \textbf{96.60} & \textbf{94.67} & \textbf{95.62} \\
\midrule
\end{tabular}
\label{tab:ablation_backbone}
\end{table*}

\subsection{\rqone}

Table~\ref{tab:jit_ccibench_full} shows the performance of three baseline models, three LLMs and our approach on two benchmarks \jitdata and \ccibench.
It can be observed that \tool consistently outperforms the others across all evaluation metrics.
Specifically, on \jitdata it reaches 90.89\% F1 on the full test set and 93.56\% on the validated subset, achieving an improvement of 5.45\%, 6.63\% and 10.90\% comparing to \texttt{CCISolver}~\cite{Zhong2025CCISolverED}, \texttt{CR4LLama}~\cite{Rong2025CodeCI} and \texttt{DeepJIT}~\cite{Panthaplackel2020DeepJI}. 
Although all three baselines encode code diffs with action labels, \tool goes further by also separating tokens into old/new spans rather than mixing all edits.
This enables the model to focus on specific modifications (additions and deletions) and highlight the mismatched semantics of inconsistent comments.
The results achieved by \tool on \ccibench are consistent but exhibit considerable improvement.
As a noise-removal subset of \jitdata, all the models achieved higher performance on \ccibench.
As can be seen in Table~\ref{tab:jit_ccibench_full}, \tool 
attains 93.38\% F1 on the full test and 95.62\% on the validated subset, exceeding \ccisolver by 4.29\% and 3.71\%, gaining an increasing of 5.31\% and 5.51\% comparing to \texttt{C4RLLama}.

We then evaluate recent LLMs by fine-tuning them for CCI detection on the same code-diff/comment training data.
The results show that \tool, built on a small model (\texttt{CodeT5+}, 110M parameters), outperforms all LLMs tested, including \deepseekcoder~(7B), \qwen~(7B), and \codellama~(7B) on both benchmark datasets.
For example, considering F1-Score, \tool achieves an improvement of 11.13\%, 6.19\% and 10.48\% on \jitdata and an increasing of 10.54\%, 5.50\% and 9.70\% on \ccibench, comparing to \codellama, \qwen and \deepseekcoder, respectively. 
Among the LLMs, \qwen performs best, reaching nearly 87\% F1 and indicating substantial promise over current LLM baselines.

\vspace{.2cm}
\begin{tcolorbox}[boxrule=0.86pt,left=0.3em, right=0.3em,top=0.1em, bottom=0.05em, ]
	\small{\textbf{Answer to RQ$_1$.} \tool demonstrates superior performance across both datasets, surpassing both prior works and recent fine-tuned LLMs, establishing itself as the leading approach for Just-in-time CCI detection.}
\end{tcolorbox}

\subsection{\rqtwo}
We examine three ablation study configurations: (i) using \tool without activity labeling and contrastive loss; (ii) using \tool without activity labeling; and (iii) using \tool without contrastive loss.
As can be seen in Table~\ref{tab:ablation_backbone}, eliminating both components (activity labeling and the contrastive loss) produces the lowest performance. Compared with the full model on the full test set, F1 decreases by 3.92\% on \jitdata and 4.37\% on \ccibench when omitting both components.
Incorporating only contrastive loss (\tool$_{\text{w/o AL}}$) slightly improves the performance. Regarding the F1-Score on \ccibench, the model achieves an increasing of 0.85\% on the full test set and 1.88\% on the validated test set comparing to the lowest configuration.
However, when we omit the contrastive loss and retain only activity labeling (\tool$_{\text{w/o CL}}$), performance remains comparable to the full configuration across both datasets and all metrics. For instance, on the full test set, F1 improves by 4.01\% on \ccibench and 3.75\% on \jitdata relative to the lowest-performing setup. A similar pattern holds on the validated set, with gains of 5.11\% on \ccibench and 4.06\% on \jitdata.
This demonstrates that the decomposition of code-diff sequence into \texttt{old} and \texttt{new} token spans enables the model to more effectively detect inconsistent comments.
Overall, the results validate that both modules complement each other to achieve robust {\it just-in-time} inconsistency detection. 


\vspace{.2cm}
\begin{tcolorbox}[boxrule=0.86pt,left=0.3em, right=0.3em,top=0.1em, bottom=0.05em]
	\small{\textbf{Answer to RQ$_2$.} 
    The most consequential factor is the explicit decomposition of code diffs into old and new modifications; other component, the label-aware contrastive loss further reinforce performance.
   }
\end{tcolorbox}

\subsection{\noindent{Threats to Validity}}

\subsubsection{Internal Validity}
Our approach is based on the assumption that each code–comment pair before a code change is consistent. 
If this assumption does not hold, incorrect supervision signals may be introduced during training. We evaluated the proposed approach and other baselines on \jitdata and its noise-removal version, \ccibench. 
Although \ccibench has been carefully cleaned to remove most mislabeled samples, a few noisy records may still remain, as some parts of the cleaning process involved LLMs, which are not fully error-free~\cite{Zhong2025CCISolverED}.

\subsubsection{External Validity}
One threat to external validity is that our evaluation is limited to benchmark datasets with about 33,000 training examples—small relative to the broader code–comment space. In addition, the data covers only Java projects, which may constrain generalizability. Nonetheless, because Java is widely used in code–comment research, we contend that the core principles of our approach readily transfer to other programming languages.

	
	
	\section{Conclusion and Future Work}
	\label{sec:conclusion}
	In this paper, we present \tool for automated just-in-time code–comment inconsistency detection. Our method combines a decomposition of code diffs into activity-labeled spans with label-aware contrastive learning in a bimodal alignment framework. Experiments on \jitdata and \ccibench show that \tool outperforms recent baselines and representative fine-tuned LLMs, underscoring the benefits of pairing effective components with a compact model.
For future work, we will pursue larger and more diverse datasets, study scalability across programming languages, and extend the approach to multilingual code and more complex real-world code–comment settings.



\section*{Acknowledgment} 
This paper has been partially supported by the MOSAICO project (Management, Orchestration and Supervision of AI-agent COmmunities for reliable AI in software engineering) that has received funding from the European Union under the Horizon Research and Innovation Action (Grant Agreement No. 101189664). 
The work has been also partially supported by the European Union--NextGenerationEU through the Italian Ministry of University and Research, Projects PRIN 2022 PNRR \emph{``FRINGE: context-aware FaiRness engineerING in complex software systEms''} grant n. P2022553SL. We acknowledge the Italian ``PRIN 2022'' project TRex-SE: \emph{``Trustworthy Recommenders for Software Engineers,''} grant n. 2022LKJWHC. 



\bibliographystyle{IEEEtran}
\bibliography{IEEEabrv,main}

\end{document}